\documentclass[pdflatex,sn-mathphys-num]{sn-jnl}

\usepackage{graphicx}%
\usepackage{multirow}%
\usepackage{amsmath,amssymb,amsfonts}%
\usepackage{amsthm}%
\usepackage{mathrsfs}%
\usepackage[title]{appendix}%
\usepackage{xcolor}%
\usepackage{textcomp}%
\usepackage{manyfoot}%
\usepackage{booktabs}%
\usepackage{algorithm}%
\usepackage{algorithmicx}%
\usepackage{algpseudocode}%
\usepackage{listings}%

\theoremstyle{thmstyleone}%

\theoremstyle{thmstyletwo}%

\theoremstyle{thmstylethree}%

\begin{document}

\title[Article Title]{On the Generalization Limits of Quantum Generative Adversarial Networks with Pure State Generators}

\author[1]{\fnm{Jasmin} \sur{Frkatovic}}

\author[1,4]{\fnm{Akash} \sur{Malemath}}

\author[1]{\fnm{Ivan} \sur{Kankeu}}

\author[1,2]{\fnm{Yannick} \sur{Werner}}

\author[1]{\fnm{Matthias} \sur{Tsch\"ope}}

\author[1,2]{\fnm{Vitor} \sur{Fortes Rey}}

\author[3]{\fnm{Sungho} \sur{Suh}}

\author[1,2]{\fnm{Paul} \sur{Lukowicz}}

\author[1,2]{\fnm{Nikolaos} \sur{Palaiodimopoulos}}

\author[1,2,4]{\fnm{Maximilian} \sur{Kiefer-Emmanouilidis}}

\affil[1]{\orgdiv{Department of Computer Science and Research Initiative QC-AI}, \orgname{RPTU Kaiserslautern-Landau}, \orgaddress{\city{Kaiserslautern}, \postcode{67663}, \country{Germany}}}

\affil[2]{\orgdiv{Embedded Intelligence}, \orgname{German Research Center for Artificial Intelligence (DFKI)}, \orgaddress{\city{Kaiserslautern}, \postcode{67663}, \country{Germany}}}

\affil[3]{\orgdiv{Department of Artificial Intelligence}, \orgname{Korea University}, \orgaddress{\city{Seoul}, \postcode{20841}, \country{Republic of Korea}}}

\affil[4]{\orgdiv{Department of Physics}, \orgname{RPTU Kaiserslautern-Landau}, \orgaddress{\city{Kaiserslautern}, \postcode{67663}, \country{Germany}}}

\abstract{We investigate the capabilities of Quantum Generative Adversarial Networks (QGANs) in image generations tasks. Our analysis centers on fully quantum implementations of both the generator and discriminator. Through extensive numerical testing of current main architectures, we find that QGANs struggle to generalize across datasets, converging on merely the average representation of the training data. When the output of the generator is a pure-state, we analytically derive a lower bound for the discriminator quality given by the fidelity between the pure-state output of the generator and the target data distribution, thereby providing a theoretical explanation for the limitations observed in current models. Our findings reveal fundamental challenges in the generalization capabilities of existing quantum generative models. While our analysis focuses on QGANs, the results carry broader implications for the performance of related quantum generative models.}


\maketitle

\section{Introduction}\label{sec1}

Over the past decade, advancements in model architectures, the availability of larger datasets, and improvements in hardware—among other factors—have significantly enhanced the capabilities of generative machine learning models \cite{harshvardhan2020comprehensive,kaplan2020scaling,sarker2021deep}. At the same time, ongoing progress toward scalable quantum hardware has sparked growing interest in the development of quantum machine learning (QML) algorithms \cite{Wang_2024, gujju2024quantum}, which aim to leverage quantum properties—such as superposition and entanglement—to enhance the efficiency and expressivity of classical machine learning approaches. Although large-scale fault-tolerant quantum hardware is not yet realizable, many QML algorithms are specifically designed to operate within the constraints of the noisy intermediate-scale quantum (NISQ) era \cite{bharti2022noisy,coyle2022machine,meyer2021fisher}. Within this broader context, QML has been applied to quantum-state analysis, including entanglement characterization, as well as to variational models for quantum and classical data, highlighting its growing scope in near-term quantum systems. \cite{chen2021detecting,ma2018transforming,liu2021rigorous,xiao2023practical,xu2025toward}.

In image generation tasks, several classical deep learning architectures have demonstrated notable effectiveness. Variational Autoencoders (VAEs) are particularly useful for tasks like image denoising \cite{prakash2020fully} and anomaly detection \cite{an2015variational} due to their structured latent spaces. Generative Adversarial Networks (GANs) \cite{goodfellow2014generative} have excelled in realistic image synthesis \cite{karras2021alias,sauer2023stylegan} and image-to-image translation \cite{choi2018stargan,richardson2021encoding}, benefiting from their adversarial training mechanism. More recently, diffusion models have emerged as state-of-the-art for tasks such as text-to-image generation \cite{gu2022vector} and high-resolution image synthesis \cite{rombach2022high}, offering improved output quality and diversity. 

Quantum analogues of the aforementioned architectures have been proposed \cite{khoshaman2018quantum,lloyd2018quantum,parigi2024quantum}; however, their efficiency currently lags behind that of their classical counterparts. In particular, it is still unclear whether fully quantum generative models can genuinely learn complex data distributions rather than reproducing dominant features of the training data. In this work, we therefore focus on Quantum Generative Adversarial Networks (QGANs), which have reported the most promising results for quantum image generation to date.

Classical GANs consist of two competing neural networks—the generator and the discriminator—engaged in a dynamic min-max game. The generator attempts to produce data that mimics the real distribution, while the discriminator aims to distinguish between real and generated data. This adversarial training continues until a Nash equilibrium is reached, ideally when the generator produces outputs indistinguishable from real data. The first proposals toward the development of Quantum Generative Adversarial Networks (QGANs) were introduced in Refs. \cite{lloyd2018quantum,dallaire2018quantum}, where the potential advantages of quantum models were demonstrated, while Ref. \cite{dallaire2018quantum} specifically showed that training is feasible within the quantum framework using a simple circuit. Subsequent studies introduced hybrid QGANs, in which one component—typically the discriminator—is implemented as a classical network, while the other remains quantum. These hybrid models have been applied to learning discrete \cite{situ2020quantum,Riofrio2024}, continuous \cite{romero2021variational, Riofrio2024}, and random distributions \cite{zoufal2019quantum}. Image generation tasks have been addressed using hybrid QGANs through various techniques, including a patching scheme \cite{huang2021experimental,li2021quantum}, learning a discrete distribution \cite{zhou2023hybrid}, employing a quantum Wasserstein GAN architecture \cite{tsang2023hybrid}, and replacing the patching and principal component analysis by a quantum long-short term memory approach \cite{chu2025lstmqganscalablenisqgenerative}. Alongside hybrid approaches, several fully quantum GAN schemes have been proposed. These include entangling QGANs \cite{niu2022entangling}, approaches leveraging quantum optimal control techniques \cite{kim2024hamiltonian}, and architectures based on the Wasserstein GAN framework \cite{chakrabarti2019quantum}. Discrete datasets have also been explored in this context \cite{chaudhary2023towards}, and some of these models have been experimentally implemented on superconducting circuits \cite{hu2019quantum, huang2021quantum}. For image generation tasks QuGAN \cite{stein2021qugan} and IQGAN \cite{chu2023iqgan} have shown the most promising results. Similar architectures like EQ-GAN \cite{niu2022entangling} have been used for efficient state preparations, but in principle could be employed for image generation as well.

In this work, we focus on the QuGAN and IQGAN models and evaluate their generalization capabilities. Specifically, we investigate whether their reported performance reflects genuine distribution learning or instead convergence to dominant features of the training data. Our numerical findings motivate a closer analysis of the generator’s role within the QGAN framework, ultimately revealing fundamental limitations that arise when the generator outputs a single pure quantum state. 

The paper is organized as follows: In section \hyperref[sec:eval]{"QGANs for Image Generation"} we evaluate the performance of the state-of-the-art QGAN architectures which is followed by an analytic derivation of a lower bound explaining the failure of generalization of current architectures in section \hyperref[sec:proof]{"Fidelity Bounds in QGANs"}. In section \hyperref[sec:conc]{"Conclusions"} we conclude our results. In the Appendices we show additional information such as the loss curves and numerical examples, how noise affects a quantum generator.

\section{QGANs for Image Generation} 
\label{sec:eval}
In this section, we begin by evaluating the two fully quantum GAN models that have reported promising results in image generation: QuGAN \cite{stein2021qugan} and IQGAN \cite{chu2023iqgan}. Both models are trained on the MNIST dataset of handwritten digits \cite{scikit-learn-digits-classification}. Our analysis starts with IQGAN, the most recent of the two, followed by a detailed examination of QuGAN. We apply a series of numerical tests aimed at highlighting the challenges encountered during training, underscoring the urgent need for more rigorous and standardized testing frameworks \cite{bowles2024betterclassicalsubtleart}.

\begin{figure}[h]
\centering
\includegraphics[width=0.9\textwidth]{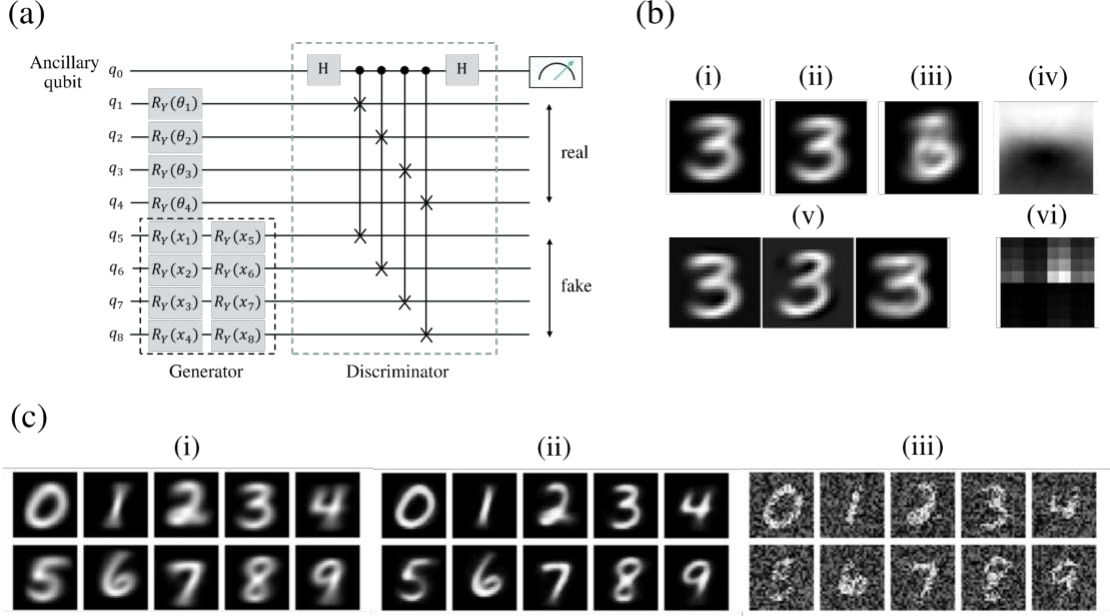}
\caption{(a) IQGAN circuit. (b) (i)  Generated image after training the IQGAN circuit on class 3 of the MNIST dataset, using PCA, an 8-qubit circuit and angle embedding. (ii) Average image of MNIST class 3. (iii) Generated image after simultaneously training the IQGAN on MNIST classes 3 and 6. (iv) Generated image after training on class 0 of the CIFAR-10 dataset, with PCA, an 8-qubit circuit and angle embedding. (v) Images produced by the inverse PCA model, using a random input (range [0,1] and normalized). (vi) Generated image after training the IQGAN on class 3 of MNIST without PCA, using a 16-qubit circuit, reducing the size of the training samples by 50\% and using amplitude embedding. (c) (i) Average over each class of MNIST. (ii) Generated images of IQGAN-784 for each class without embedded noise. (iii) Generated images of IQGAN-784 for each class with embedded noise.}\label{fig1}
\end{figure}

In most QGAN architectures, both the generator and the discriminator are typically implemented as Quantum Variational Circuits (QVCs). The IQGAN architecture (see Fig. \ref{fig1}(a)) though works a bit differently. This circuit employs angle embedding via a fixed or trainable encoder. The lower part of the circuit corresponds to the generator, which is a trainable QVC that generates fake data. While in the upper part real data are encoded and the discriminator is essentially a swap test \cite{buhrman2001quantum,schuld2021machine} comparing the two.

At this point, it is worth clarifying the distinction between adversarial learning and a Quantum Circuit Born Machine (QCBM). A QCBM optimizes a single parametrized quantum circuit using a direct, often fidelity-based, objective rather than an adversarial discriminator. As a result, the generator is trained in isolation instead of through a min–max game, leading to fundamentally different training dynamics. This distinction becomes relevant when interpreting the behavior of IQGAN below.

We highlight several key aspects of this approach that, as we will see, are crucial to its performance. First, the model is trained separately on each class of the MNIST dataset. Second, a Principal Component Analysis (PCA) pre-processing step is applied before encoding, which drastically reduces the dimensionality of each training image—from 784 pixels down to just 4. This represents an extreme compression of the data. Additionally, the swap test, which functions as the discriminator in this setup, is not a trainable quantum variational circuit (QVC). By definition, this makes the discriminator non-trainable, distinguishing it from typical adversarial learning frameworks. The swap test therefore induces a fixed, fidelity-based cost function rather than a learnable adversarial objective. As a result, the IQGAN circuit aligns more closely with a Quantum Circuit Born Machine (QCBM)\cite{liu2018differentiable}, acting as an optimized generator without engaging in a min-max adversarial game. Finally, we know that in classical GAN architectures, the generator receives a noise vector as input, enabling the production of varied outputs that resemble the distribution of the training data \cite{goodfellow2014generative,goodfellow2020generative}. In contrast, IQGAN does not incorporate any input noise, which may limit its capacity to generate diverse outputs.

Let us now proceed with the evaluation of the model. In Fig.~\ref{fig1}(b)(i), we present the image generated by the model after being trained on class 3 of the MNIST dataset. Note, that the corresponding loss curve is given in the ``Supplementary Material”. Right next to it in Fig.~\ref{fig1}(b)(ii), we show the class average for comparison. As can be seen, the two images are nearly identical. This unexpected similarity raised concerns about the actual generative capabilities of the QVC, prompting us to consider whether the PCA step was solely responsible for the observed output. To investigate this further, we replaced the QVC with a random number generator. Surprisingly, the resulting outputs (see Fig.~\ref{fig1}(b)(v)) were visually similar—suggesting that the quantum circuit is not contributing to the generation process in its current form.

To verify this, we conducted additional tests. First, we trained the model on two classes of the MNIST dataset simultaneously (see Fig.~\ref{fig1}(b)(iii)). Second, we evaluated the model on CIFAR-10 \cite{krizhevsky2009learning}— a more complex and diverse dataset (see Fig.~\ref{fig1}(b)(iv)). In both cases, the model failed to generate meaningful results, suggesting that it primarily captures an average representation of the training data rather than learning its underlying distribution, which in general is a problem of current QNN \cite{werner2025disquinvestigatingimpactdisorder}.

Next, we conducted two experiments in which the PCA pre-processing step was entirely removed. In the first test, the images were downsampled by 50$\%$, resulting in 14×14 resolution, and amplitude encoding was applied using 16 qubits. As shown in Fig.~\ref{fig1}(b)(v), the model exhibits behaviour resembling mode collapse.

Moreover, the best and most gate efficient reported IQGAN circuit in \cite{chu2023iqgan} was missing all entangling gates. Thus, we conducted an additional experiment in which we scaled the circuit to 784 qubits—one for each pixel—treating them independently during training. In this configuration the embedding circuit (angle encoding) and the generator are equivalent. Thus, the generator only learns the corresponding angles that correspond to the image embedding. Learning a set of angles where each angle corresponds to a single pixel value can effectively interpreted as a look-up table and does not generalize. In Fig.~\ref{fig1}(c)(i), we display the average image for each class. Next to it, in Fig.~\ref{fig1}(c)(ii), we present the images generated by the model which closely resemble the average of each class, further supporting the idea that the model fails to learn the real data distribution. In (see In Fig.~\ref{fig1}(c)(iii)) we show the generated outputs but this time using embedded Gaussian noise as input. As shown, the model performs poorly, underscoring its lack of generalization under such conditions.

These findings lead us to conclude that the model's generative behaviour is heavily dependent on the PCA reduction step, and that, in practice, it functions primarily as a mechanism for reproducing the average of the dataset.

The methodology used in the QuGAN approach is similar to that of IQGAN, with the primary difference lying in the design of the quantum circuits. As in IQGAN, a PCA pre-processing step is applied to reduce the input dimensionality, followed by angle embedding; however, in this case, a fixed encoder is used. 

\begin{figure}[h]
\centering
\includegraphics[width=0.8\textwidth]{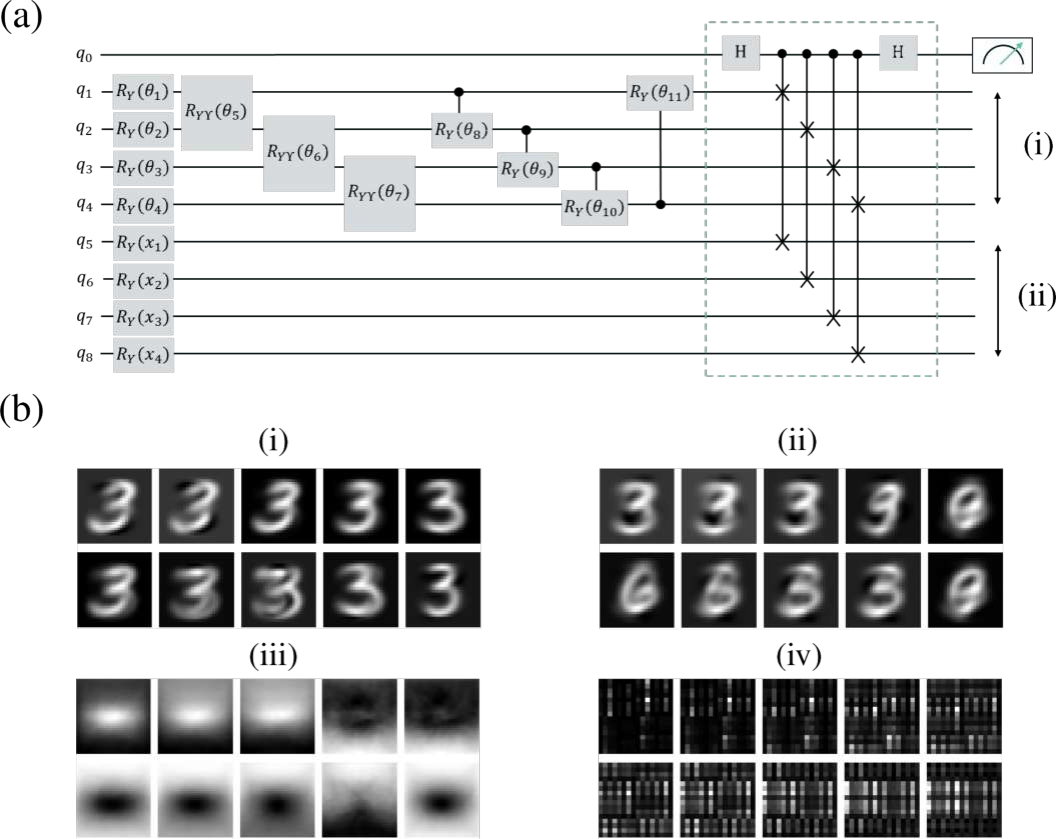}
\caption{(a) QuGAN circuit architecture. (b) Generated images under various training settings: (i) QuGAN trained on class 3 of MNIST using PCA, an 8-qubit circuit, and angle embedding. (ii) QuGAN trained simultaneously on classes 3, 6, and 9 of MNIST with PCA, an 8-qubit circuit, and angle embedding. (iii) QuGAN trained on class 0 of CIFAR-10 using PCA, an 8-qubit circuit, and angle embedding. (iv) QuGAN trained on class 3 of MNIST without PCA, using a 16-qubit circuit, 50\% downsampling of training data, and amplitude embedding. Note, that each generated image was sampled every 40 batches during the training process.}\label{fig2}
\end{figure}

In Fig.~\ref{fig2}(a) we show the circuit of QuGAN. As we will explain, both the discriminator and the generator are implemented as QVCs. The QuGAN approach feature a two-step training process. In the first step, in the upper part of the circuit (Fig.~\ref{fig2}(a)(i)) a QVC serves as the discriminator. Fake data is passed through this circuit, while real data is encoded in the lower part of the circuit (Fig.~\ref{fig2}(a)(ii)). A swap test, similar to that used in the IQGAN architecture, is performed to distinguish between real and generated data. In the second step of training, the lower part of the circuit is replaced with the generator’s QVC, which is architecturally identical to that of the discriminator—ensuring fairness in the training setup.

The circuit design of both the discriminator and generator consists of three layers. The first layer applies single-qubit unitary operations using $R_Y$ rotation operations. The second layer introduces two-qubit interactions via $IsingYY$ gates. Finally, the third layer establishes entanglement across qubits using controlled  $CR_Y$ operations.

Unlike IQGAN, QuGAN trains multiple classes simultaneously, specifically classes 3, 6, and 9. This leads to performance drop compared to IQGAN, primarily due to the increased complexity introduced by multi-class training. 

Since the approaches employed by QuGAN and IQGAN are fundamentally similar, we expect their results to be comparable as well. Indeed, upon examining Figs.~\ref{fig2}(b)(i) and (ii)—which show the generated outputs when the model is trained on class 3 and on classes 3, 6, and 9 of MNIST, respectively—we observe similar outcomes. The results produced by QuGAN appear slightly inferior; however, as discussed above, this is expected due to the increased complexity introduced by multi-class training. In this case, it is evident from Fig.~\ref{fig2}(b)(ii) that the generated images transition between classes during training. This oscillatory behaviour, although subtle, indicates that the model has difficulty converging to a stable solution. This is also evident when inspecting the corresponding loss curves for Figs.~\ref{fig2}(b)(i) and (ii) which are given in Appendix \ref{secA1}.

The generated images reflect a model that attempts to learn the average characteristics of the training data, yet ultimately fails—likely due to the unstable adversarial interplay between the QVCs. When examining the best generated outputs, which clearly resemble a single digit, it becomes apparent that these images represent the average of that specific class. This observation is supported by the results shown in Fig.~\ref{fig2}(b)(i), where the model was trained exclusively on class 3 of the MNIST dataset. If the training process were to stabilize, the resulting images would likely be indistinguishable from those produced by IQGAN. Moreover, transforming QuGAN into a Quantum Circuit Born Machine (QCBM) by removing the discriminator would produce similar outputs. However, this modification would eliminate the core adversarial mechanism that defines a QGAN.

In Fig.~\ref{fig2}(b)(iii) we show the generated output of QuGAN when trained to class 0 of CIFAR-10. As expected, the results closely resemble that of IQGAN, since both models heavily rely on PCA. The last thing we tested is whether the larger circuit of QuGAN (incorporating more trainable parameters) could enhance its performance when we omit the steps involving the PCA. To this end, we doubled the circuit size from 8 to 16 qubits, employed amplitude embedding and downsampled by 50\% the images of the training dataset. The generated outputs are depicted in Fig.~\ref{fig2}(b)(iv) and closely resembling noise—similar to the IQGAN results. We note, that this problem has also been addressed by \cite{chu2025lstmqganscalablenisqgenerative}.

To complement the qualitative analysis presented above, we now turn to a quantitative evaluation of the generated samples using standard distribution-level metrics \cite{silver2023mosaiq,ma2025quantum}. Specifically, we assess the similarity between real and quantum-generated images by computing the Fréchet Inception Distance \cite{heusel2017gans} (FID) on MNIST (digit 3) and CIFAR-10 (class 0), which we select as representative benchmark classes. After training, the quantum generator is sampled by performing projective measurements on the generator qubits in the computational basis, yielding binary bitstrings that serve as latent representations. For example, with four generator qubits, a single measurement may produce a bitstring such as \(0101\), corresponding to a four-dimensional latent vector. Repeating this procedure yields an empirical distribution over the corresponding latent representations.

These bitstrings are mapped back to the continuous data space by applying the inverse MinMax normalization followed by the inverse principal component analysis (PCA), resulting in reconstructed grayscale images. Both real and generated images are resized to \(299 \times 299\), replicated across three channels, and processed using a pretrained Inception-v3 network with its classification head removed to extract 2048-dimensional feature representations. The mean \(\mu\) and covariance \(\Sigma\) of these features are computed for each dataset, and the FID is defined as
\begin{equation*}
\mathrm{FID} = \|\mu_{\mathrm{real}} - \mu_{\mathrm{gen}}\|_2^2
+ \mathrm{Tr}\!\left(
\Sigma_{\mathrm{real}} + \Sigma_{\mathrm{gen}}
- 2\bigl(\Sigma_{\mathrm{real}}\Sigma_{\mathrm{gen}}\bigr)^{1/2}
\right),
\end{equation*}
where lower values indicate greater similarity between the real and generated data distributions.

In addition to samples obtained from the trained quantum generators, we evaluate a classical reference baseline based on uniformly random sampling in the latent space. Specifically, for a fixed number of generator qubits, we sample binary bitstrings uniformly at random, corresponding to projective measurements of an unstructured generator that assigns equal probability to all computational basis states. These randomly sampled bitstrings are treated identically to the quantum-generated samples. They are mapped to continuous latent vectors via inverse MinMax normalization, transformed back to the data space using the inverse PCA, and post-processed into images that are subsequently used for FID evaluation. This random-sampling baseline represents the absence of any learned structure in the latent distribution and provides a reference for evaluating whether adversarial training improves the quality of the generated samples.

\begin{figure*}[h]
\centering
\includegraphics[width=0.8\textwidth]{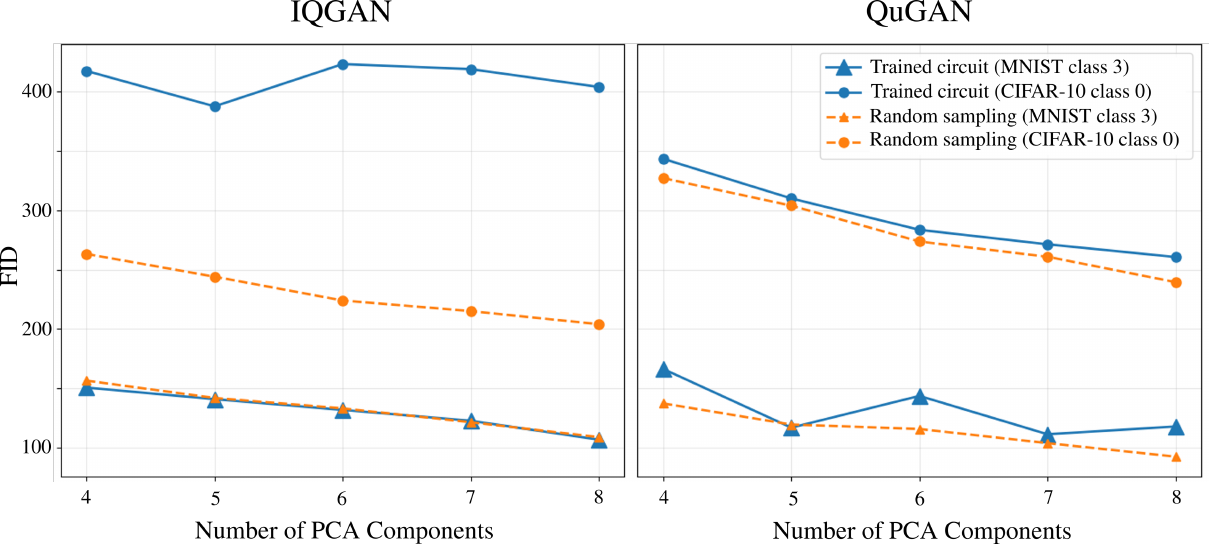}
\caption{FID values as a function of the number of principal components. For each model, samples drawn from the trained generator circuit (blue markers with solid lines) are compared against a classical baseline obtained by uniformly random sampling of latent bitstrings (orange markers with dashed lines).}
\label{fig:FID}
\end{figure*}

As shown in Fig.~\ref{fig:FID}, across different numbers of PCA components, both IQGAN and QUGAN achieve FID values that are comparable to those obtained from the uniformly random baseline. This behavior indicates that, within the considered regimes, adversarial training fails to induce a meaningful latent distribution that improves upon unstructured random sampling. The near overlap between the trained-generator curves and the random baseline thus provides clear evidence of the limited expressive power of the investigated quantum circuits at the tested system sizes and circuit depths. Notably, in the IQGAN setting applied to the more complex CIFAR-10 dataset, we observe regimes in which uniform random sampling yields lower FID values than sampling from the trained quantum generator, further underscoring that adversarial training does not reliably improve sample quality.

We have observed that both models exhibit poor generalization performance, despite employing well-motivated architectures and theoretically grounded design choices. A crucial point we wish to emphasize is that, in both IQGAN and QuGAN, there is no sampling over latent variables, no classical mixing over circuit parameters, and no stochastic measurement-induced distribution. Consequently, the generator represents the entire model distribution by a single deterministic quantum state. To ensure reproducibility, in Appendix \ref{secA2} we provide information on the training procedures and hyperparameters used in our numerical simulations. Motivated by the aforementioned observation, in the next section, we shed light on the underlying reasons for this limitation by deriving analytical bounds on the fidelity achieved by QGANs.

\section{Fidelity bounds in QGANs}
\label{sec:proof}
We assume a QGAN that performs equivalently to its classical counterpart, evaluated using a fidelity-based metric. It is therefore comprised of a parametrized quantum circuit acting as a generator $G(z)$ and a quantum discriminator $D(x)$. The generator maps samples from a given input distribution to data examples, thereby creating fake data. Training is formulated as a min-max optimization problem that balances the discriminator’s ability to distinguish real data from the generator’s output against the generator’s ability to fool the discriminator. Accordingly, the expectation value of the discriminator  $D$ assigning high probability to real data is contrasted with the expectation value of the generator $G$ producing samples from noise $z$, drawn from a given probability distribution $p(z)$, that successfully deceive the discriminator:
\begin{equation}\label{eq:qgn}
\min_G \max_D V(D, G) =\min_G \max_D \mathbb{E}_{x \sim p_{\text{data}}}[\log D(x)] + \mathbb{E}_{z \sim p_z}[\log(1 - D(G(z)))].
\end{equation}
For fixed generator state $\rho_G$ and data state $\rho_{\text{data}}$, the optimal minimum-error binary measurement for discriminating between the two states, given prior probabilities $\pi_{\text{data}}$ and $\pi_G$, attains the Helstrom success probability \cite{helstrom1969quantum,barnett2009quantum}
\begin{equation}\label{eq:helstrom}
P_{\mathrm{succ}}^{\star}
= \frac{1}{2}\left(1 + \left\| \pi_{\text{data}}\rho_{\text{data}} - \pi_G\rho_G \right\|_1 \right),
\end{equation}
where $\|\cdot\|_1$ denotes the trace norm and $\pi_G = 1-\pi_{\text{data}}$. In the balanced setting $\pi_{\text{data}}=\pi_G=\tfrac12$ this becomes
$P_{\mathrm{succ}}^{\star}=\tfrac12\left(1+\tfrac12\|\rho_{\text{data}}-\rho_G\|_1\right)$.

Identifying the discriminator's optimal average output with this success probability, i.e. $D^\star=D(x)=D(G(z)) := P_{\mathrm{succ}}^{\star}$,
we obtain for the value function
\begin{equation}
\begin{aligned}
V(D^\star,G)
&= \log\!\left(\frac{1}{2}+\frac{1}{4}\left\|\rho_{\text{data}}-\rho_G\right\|_1\right)
 + \log\!\left(\frac{1}{2}-\frac{1}{4}\left\|\rho_{\text{data}}-\rho_G\right\|_1\right).
\end{aligned}
\end{equation}
In particular, at $\rho_{\text{data}}=\rho_G$ we have $\|\rho_{\text{data}}-\rho_G\|_1=0$ and therefore
\begin{equation}
V(D^\star,G) = \log\!\left(\tfrac12\right)+\log\!\left(\tfrac12\right) = -2\log 2.
\end{equation}

More generally the existence of a Nash equilibrium at the desired location in the QGAN setting has been shown by \cite{niu2022entangling}.
This matches the classical GAN case, where the optimum is reached when \( p_{\text{data}} = p_G \), and the optimal discriminator is:
\begin{equation}
D_{\mathrm{cl}}^*(x) = \frac{p_{\text{data}}(x)}{p_{\text{data}}(x) + p_G(x)}.
\end{equation}

In the quantum setting, the discriminator corresponds to a fixed measurement optimized to distinguish whether an input state originates from the data ensemble $\rho_{\text{data}}$ or from the generator ensemble $\rho_G$. For given $\rho_{\text{data}}$ and $\rho_G$, the optimal strategy is the Helstrom measurement \cite{helstrom1969quantum,barnett2009quantum}, defined by the positive and negative eigenspaces of the operator
$\pi_{\text{data}}\rho_{\text{data}} - \pi_G\rho_G$.
In the balanced case, this measurement depends only on the difference $\rho_{\text{data}} - \rho_G$. For concreteness, and without loss of generality in the pure-state case, the Helstrom measurement may be represented as a projective measurement onto the eigenbasis of this operator. Let $|\delta\rangle$ denote a normalized eigenstate associated with the positive eigenspace. The discriminator output probabilities for real and generated inputs are then
\begin{align}
\begin{split}
D(x_i) &= |\langle \delta | x_i \rangle|^2, \\
D(G(z)) &= |\langle \delta | \gamma \rangle|^2,
\end{split}
\end{align}
where $|x_i\rangle = E_i |0\rangle$ denotes an encoded real data sample drawn from the ensemble
$\rho_{\text{data}} = \sum_i p_i |x_i\rangle\langle x_i|$,
and $|\gamma\rangle = G|0\rangle$ is the generator output corresponding to latent input $z$. Although the measurement outcome probabilities depend on the particular input state, the discriminator itself is fully specified by a single fixed measurement determined solely by $\rho_{\text{data}}$ and $\rho_G$, and does not adapt to individual samples. In this sense, the quantum discriminator is directly analogous to the classical optimal discriminator, which is completely determined by the underlying data and generator distributions.

The distinguishability of $\rho_{\text{data}}$ and $\rho_G$ appearing in the Helstrom bound can be related to the fidelity via the Fuchs--van de Graaf inequalities \cite{fuchs2002cryptographic},
\begin{equation}
1 - F(\rho_{\text{data}}, \rho_G)
\leq \frac{1}{2}\left\| \rho_{\text{data}} - \rho_G \right\|_1
\leq \sqrt{1 - F(\rho_{\text{data}}, \rho_G)^2},
\end{equation}
providing a direct connection between optimal discrimination performance and fidelity-based metrics commonly used in quantum generative modeling. Applying this inequality yields
\begin{equation}\label{eq:fidin}
\frac{1}{2} + \frac{1}{4}\left\| \rho_{\text{data}} - \rho_G \right\|_1
\ge
1 - \frac{1}{2} F(\rho_{\text{data}}, \rho_G),
\end{equation}
where by $F(\rho,\sigma) =  \operatorname{Tr} \left(\sqrt{ \sqrt{\rho}\,\sigma\,\sqrt{\rho} } \right)$ we denote the fidelity.

If the generated state is pure, $\rho_G = |\gamma\rangle\langle \gamma|$, then
\begin{equation}\label{eq:uhlmann_pure}
F(\rho_{\text{data}}, \rho_G)
= \sqrt{\langle \gamma | \rho_{\text{data}} | \gamma \rangle}.
\end{equation}
For an ensemble decomposition
$\rho_{\text{data}} = \sum_{i=1}^r p_{\text{data},i}\, |x_i\rangle\langle x_i|$,
this overlap evaluates to
\begin{equation}
\langle \gamma | \rho_{\text{data}} | \gamma \rangle
= \sum_{i=1}^r p_{\text{data},i}\, |\langle \gamma | x_i \rangle|^2.
\end{equation}
The maximum overlap is achieved by choosing $|\gamma\rangle$ to be an eigenvector of $\rho_{\text{data}}$ corresponding to its largest eigenvalue $\lambda_{\max}(\rho_{\text{data}})$, yielding
\begin{equation}
F_{\max} = \sqrt{\lambda_{\max}(\rho_{\text{data}})}.
\end{equation}
Consequently, Eq.~(\ref{eq:fidin}) implies the following best-case bound for any pure-state generator:
\begin{equation}\label{eq:bound}
\frac{1}{2} + \frac{1}{4}\left\| \rho_{\text{data}} - \rho_G \right\|_1
\ge
1 - \frac{1}{2} \sqrt{\lambda_{\max}(\rho_{\text{data}})}.
\end{equation}
The bound in Eq.~(\ref{eq:bound}) depends on the data state only through its largest
eigenvalue $\lambda_{\max}(\rho_{\text{data}})$, and is therefore governed by the
spectral concentration of the data ensemble. If $\rho_{\text{data}}$ has rank $r$,
then $\lambda_{\max}(\rho_{\text{data}})\geq 1/r$, yielding the worst--case scaling
$F_{\max}\sim 1/\sqrt{r}$ for flat spectra. For low--rank or spectrally peaked data,
$\lambda_{\max}(\rho_{\text{data}})$ is close to unity. Equality in
Eq.~(\ref{eq:bound}) is attained in the extremal cases where
$\rho_{\text{data}}=\rho_G$, corresponding to vanishing trace distance, or when
$\rho_{\text{data}}$ and $\rho_G$ have orthogonal support, in which case
$\|\rho_{\text{data}}-\rho_G\|_1=2$ and the fidelity becomes zero. For mixed data states, a pure-state generator cannot achieve vanishing trace distance, instead, the optimal solution corresponds to recovering the principal eigenstate of $\rho_{data}$, which maximizes the fidelity while remaining at finite trace distance from the full data ensemble.

\begin{figure}[h]
\centering
\includegraphics[width=0.6\textwidth]{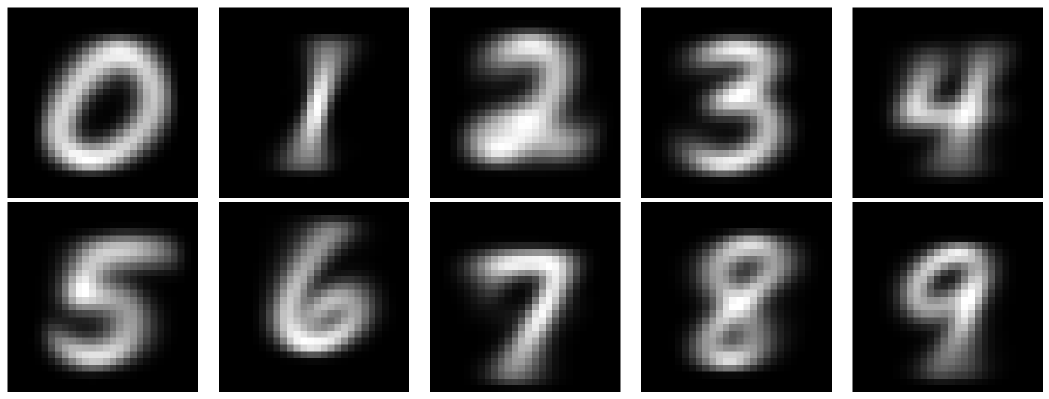}
\caption{We show the leading eigenvectors of $\rho_\text{data}$ (first principal components) for each class separately in the MNIST dataset.}\label{fig3}
\end{figure}

In Fig.~\ref{fig3}, we present the absolute value of the principal eigenstate $|x_{\max}\rangle$ of the data density operator $\rho_{\text{data}}$ for each class of the MNIST dataset, where $|x_{\max}\rangle$ denotes the eigenvector associated with the largest eigenvalue $\lambda_{\max}(\rho_{\text{data}})$. Importantly, this state does not in general coincide with any individual data sample, but instead represents a coherent superposition capturing the dominant
mode of the data ensemble. The close visual resemblance between $|x_{\max}\rangle$ and the outputs generated by QuGAN (Fig.~\ref{fig2}) and IQGAN (Fig.~\ref{fig1}) indicates that both models converge toward this principal eigenstate. This behavior is consistent with our analytical results, which show that for generators restricted to producing a single pure quantum state, the maximal achievable fidelity with $\rho_{\text{data}}$ is attained by the leading eigenvector of the data density operator. The observed convergence therefore reflects the optimal pure-state approximation of the data ensemble, rather than successful learning of the full data distribution.

The more complex (i.e., higher-rank) a dataset is, the poorer the training performance of a QGAN with a pure-state generator, due to fundamental limits on fidelity and distinguishability. This limitation extends to other quantum generative models that output a single pure quantum state. In such cases, if no sampling is performed, the task shifts from generalizing an unknown distribution to simply fitting or reproducing a known one. While this may be appropriate for applications like efficiently representing quantum states \cite{niu2022entangling} or implementing quantum look-up tables \cite{faizy2025scratchquantumcomputingreducing}, it does not constitute a generalizing approach. To this end, it is meaningful to explore the generalization capabilities of quantum generators that do not necessarily output pure states. A discussion of this direction, along with some preliminary results, is provided in Appendix \ref{secA3}.

\section{Conclusions} 
\label{sec:conc}
We have explored the generalization capabilities of two state-of-the-art QGAN models for image generation tasks. Both IQGAN and QuGAN tend to capture only the dominant average feature of the dataset, aligned with the leading principal component, and otherwise fail to perform well on datasets with high variance. This limitation stems from two primary issues: applying PCA compresses the data so aggressively that the variational quantum circuit (VQC) becomes largely unresponsive to training signals; conversely, skipping PCA reveals the generator’s limited expressivity, which prevents it from modeling richer data structures.

Beyond our numerical analysis of existing QGAN models, a central contribution of this work is the explicit analytical connection we establish between optimal discriminator performance and the eigenvalue structure of the data density matrix when the generator outputs a single pure quantum state. Namely, we analytically show that the limited expressivity of QGAN models stems from a fundamental constraint: a lower bound on the fidelity between the generator's pure-state output and the target data distribution. This bound limits the generator’s ability to match complex data, thereby conflicting with the optimal training of the quantum discriminator.

This result complements existing studies on the expressivity of quantum generative models (e.g. QGANs \cite{lloyd2018quantum,dallaire2018quantum,niu2022entangling}, PQCs \cite{sim2019expressibility,holmes2022connecting}, and QCBMs \cite{amin2018quantum}) by highlighting that approaches relying solely on pure-state outputs—without post-selection or sampling—face intrinsic limitations when modelling high-rank (i.e., mixed or complex) data distributions. In such cases, the generator cannot truly generalize; instead, it approximates a fixed representative of the data, effectively reducing the generative task to deterministic reproduction.

While our numerical simulations focus on classical datasets encoded into mixed quantum states, the analytical results presented in this paper are not specific to classical data generation and apply equally to the generation of quantum data. Our analysis is formulated entirely at the level of quantum states and quantum distinguishability measures, independent of whether the target state arises from classical data encoding or represents an intrinsically quantum dataset.

Our results highlight critical limitations in current QGAN architectures and emphasize the need for principled benchmarks and evaluation protocols to rigorously assess generalization in quantum generative learning.

\bmhead{Acknowledgements}

We gratefully acknowledge financial support from the Quantum Initiative Rhineland-Palatinate QUIP and the Research Initiative Quantum Computing for AI (QC-AI). 

\paragraph{Author Contributions} J.F., N.P., and M.K-E. mainly conducted the research. A.M., I.K., Y.W. and M.T. contributed numerical tests (not all shown) to evaluate the current feasibility of quantum generators. N.P. wrote the initial draft of the manuscript.
M.K-E. formulated the proof with help of V.F.R.. V.F.R., S.S., P.L.. N.P. and M.K-E. supervised the project. N.P. presented the results at QTML2024.
All authors contributed to the manuscript’s text and have read, reviewed and approved the final manuscript.

\paragraph{Data and Code availability}
Data and code will be made available on reasonable request to the corresponding authors N.P. and M.K-E..

\section*{Declarations}
\paragraph{Conflict of Interest} The authors declare no competing interests.

\noindent \textbf{AI Tools} like chatGPT4o, Writefull(TeXGPT), and Grammarly have been used to enhance spelling, grammar, rephrasing, as well as shortening, splitting long sentences. All improved text has been carefully read and made sure to be free of hallucinations.

\begin{appendices}\label{sec:sup]}

\section{Loss Plots} \label{secA1}
\begin{figure}[h]
\centering
\includegraphics[width=0.98\textwidth]{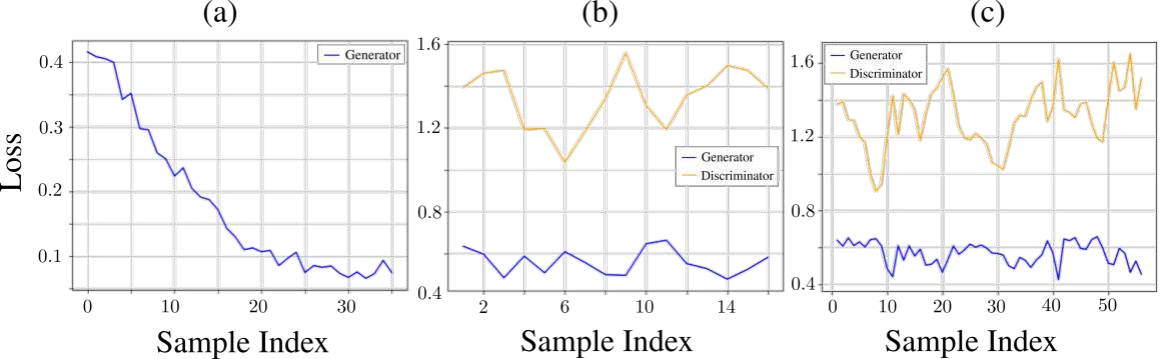}
\caption{Loss curves for: (a) IQGAN and (b) QuGAN trained on class 3 of MNIST. (c) QuGAN trained on classes 3,6, and 9 of MNIST. Loss values were sampled every 40 batches throughout the training process.}\label{figA1}
\end{figure}

In Fig.~\ref{figA1}(a) we plot the loss curve for the IQGAN model trained on class 3 of MNIST dataset. Figs.~\ref{figA1}(b) and (c) display the corresponding loss curves for the QuGAN model trained on class 3 and on classes 3, 6, and 9, respectively. We observe that the pronounced oscillations in the QuGAN model hinder stable convergence, which has been reported prior by \cite{chu2023iqgan,niu2022entangling}.

\section{Training details} \label{secA2}
All models are trained using the Adam optimizer with a learning rate of 0.001 and a batch size of 32. Training is performed for 10 epochs, which we keep fixed since the generator loss and fidelity empirically converge within this range, with no noticeable improvement beyond that. For the QGAN setup, both the generator and discriminator are parametrized quantum circuits operating on $k$ qubits, where $k$ corresponds to the number of retained PCA components. Per iteration, the discriminator is trained using a real-data minibatch and 16 fake samples generated by the generator, while the generator is subsequently updated using 8 discriminator evaluations. After each epoch, a sample is generated by measuring the generator qubits, forming a $k$-dimensional vector that is mapped back to an image via inverse normalization and inverse PCA.

\section{Limitations of a QGAN Generator}\label{secA3}

In classical GANs the generator receives noise vectors as input, enabling it to produce diverse samples from the target distribution. A key limitation that may hinder this process in variational quantum circuits (VQCs) is the absence of non-linear activation functions, a consequence of the unitary nature of quantum operations. Non-linearities are essential for enhancing the expressivity of neural networks, and their absence in quantum models could restrict the generator’s ability to model complex distributions.

\begin{figure}[h]
\centering
\includegraphics[width=0.7\textwidth]{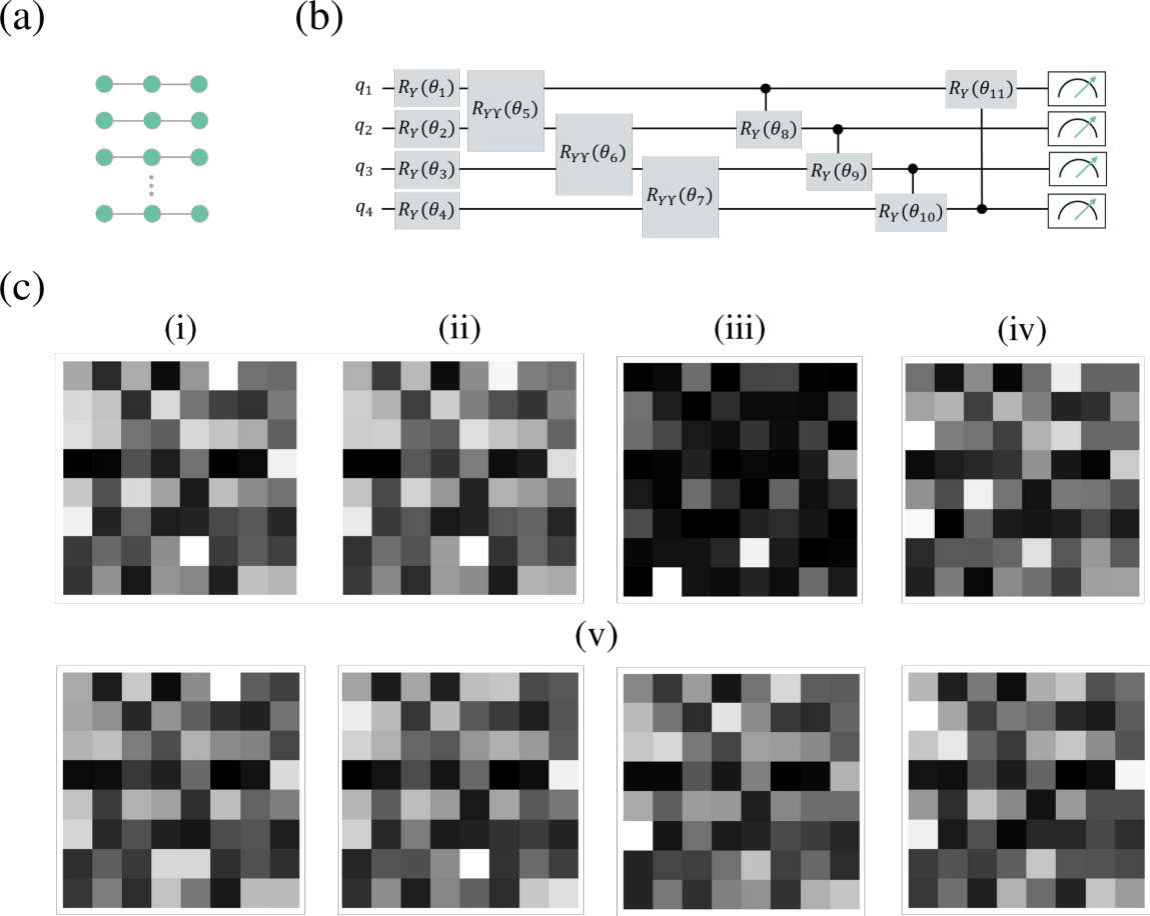}
\caption{(a) Schematic of the classical neural network architecture. (b) Quantum generator circuit used in QuGAN. (c) Visualization of target and generated images: (i) Target image of size $8 \times 8$, with pixel values sampled from a uniform distribution over [0,1]; (ii) Output image generated by the trained classical neural network; (iii) Image generated by the QuGAN generator with 6 qubits and 4 layers; (iv) Image generated by the QuGAN generator with 12 qubits, including 6 ancillary qubits; (v) Sample outputs from the trained model in (c) for various random input vectors.}\label{figA2}
\end{figure}

To highlight the importance of non-linearities, we designed a simple test comparing the effectiveness of a classical toy model of a neural network (NN) with the generator circuit used in QuGAN. In this setup, both models receive noise vectors as input at each epoch, with elements randomly sampled from a uniform distribution. The task is to reproduce a specific target image, where each pixel value is also randomly drawn from a uniform distribution.

As illustrated in the schematic of Fig.~\ref{figA2}(a), the layers of the classical NN are not fully connected. On the contrary, each neuron is connected only to its corresponding neuron in the next layer. Additionally, the weights are fixed and set to 1, ensuring that the input values remain unchanged. The only trainable parameters are the biases, which are added to the inputs before being passed through a non-linear activation function applied to the hidden layer neurons. This deliberately constrained architecture is intended to isolate and examine the effect of non-linearity. As for the QuGAN generator—one layer of which is illustrated in Fig.~\ref{figA2}(a)—we set the number of layers such that the total number of trainable parameters is approximately equal to that of the classical neural network (i.e. the number of nodes). However, we acknowledge that this does not constitute a strictly fair comparison, as the representational capacities and training dynamics of quantum and classical models differ significantly. In Fig.~\ref{figA2}(c)(i) we present the target image. Adjacent to it, Fig.~\ref{figA2}(c)(ii) shows of the classical neural network, and Fig.~\ref{figA2}(c)(iii) displays the output of the QVC. We observe that even this highly constrained version of a classical neural network is able to closely approximate the target image, whereas the QVC clearly struggles to do so. A common approach to inducing non-linear transformations in quantum circuits involves introducing ancillary qubits, performing partial measurements, and tracing out the ancillary subsystem \cite{huang2021experimental,cong2019quantum}. The non-linear effect can be amplified by increasing the number of ancillary qubits. As shown in Fig.~\ref{figA2}(c)(iv), the generated output begins to approximate the target image more closely as the number of ancillary qubits approaches that of the generator’s qubits (i.e., 12 qubits in total, with 6 ancillary). We also have to note that in this case the resulting generated state is not pure, therefore the limitations presented in Section ``Fidelity Bounds in QGANs” do not apply. In this case, once the model is trained, providing noise vectors as input leads to the generation of output states that exhibit a degree of variability (see Fig.~\ref{figA2}(c)(iv)). However, our attempts to train generative variational circuits with the aforementioned characteristics on multiple classes of the MNIST dataset were unsuccessful.

\end{appendices}

\end{document}